\documentclass[english]{article}
\usepackage[T1]{fontenc}
\usepackage[latin9]{inputenc}
\usepackage{color}
\usepackage{textcomp}
\usepackage{amstext}
\usepackage{amssymb}

\makeatletter

\newcommand{\lyxmathsym}[1]{\ifmmode\begingroup\def\b@ld{bold}
  \text{\ifx\math@version\b@ld\bfseries\fi#1}\endgroup\else#1\fi}

\makeatother

\usepackage{babel}
\begin{document}

\title{\textbf{Black hole quantum spectrum}}

\author{\textbf{Christian Corda}}

\maketitle
\begin{center}
Istituto Universitario di Ricerca \textquotedbl{}Santa Rita\textquotedbl{},
59100 Prato, Italy
\par\end{center}

\begin{center}
Institute for Theoretical Physics and Advanced Mathematics (IFM) Einstein-Galilei,
Via Santa Gonda 14, 59100 Prato, Italy
\par\end{center}

\begin{center}
International Institute for Applicable Mathematics \& Information
Sciences (IIAMIS),  Hyderabad (India) \& Udine (Italy) 
\par\end{center}

\begin{center}
\textit{E-mail address:} \textcolor{blue}{cordac.galilei@gmail.com} 
\par\end{center}
\begin{abstract}
Introducing a black hole (BH) \emph{effective temperature, }which
takes into account both the non-strictly thermal character of Hawking
radiation and the countable behavior of emissions of subsequent Hawking
quanta, we recently re-analysed BH quasi-normal modes (QNMs) and interpreted
them naturally in terms of quantum levels. In this work we improve
such an analysis removing some approximations that have been implicitly
used in our previous works and obtaining the corrected expressions
for the formulas of the horizon's area quantization and the number
of quanta of area and hence also for Bekenstein-Hawking entropy, its
sub-leading corrections and the number of micro-states, i.e. quantities
which are fundamental to realize the underlying quantum gravity theory,
like functions of the QNMs quantum ``overtone'' number $n$ and,
in turn, of the BH quantum excited level. An approximation concerning
the maximum value of $n\;$ is also corrected. On the other hand,
our previous results were strictly corrected only for scalar and gravitational
perturbations. Here we show that the discussion holds also for vector
perturbations.

The analysis is totally consistent with the general conviction that
BHs result in highly excited states representing both the ``hydrogen
atom'' and the ``quasi-thermal emission'' in quantum gravity. Our
BH model is somewhat similar to the semi-classical Bohr's model of
the structure of a hydrogen atom.

The thermal approximation of previous results in the literature is
consistent with the results in this paper. In principle, such results
could also have important implications for the BH information paradox.
\end{abstract}
\textbf{PACS NUMBERS: 04.70.Dy, 04.70.-s}


\section{Introduction}

The non-strictly thermal spectrum by Parikh and Wilczek \cite{key-1,key-2}
of Hawking radiation \cite{key-3} implies that emissions of subsequent
Hawking quanta are countable \cite{key-4,key-5,key-17,key-20,key-21,key-22,key-23,key-24},
and, in turn, generates a natural correspondence between Hawking radiation
and BH QNMs \cite{key-4,key-5,key-17}, permitting to naturally interpret
QNMs as quantum levels \cite{key-4,key-5,key-17}. In fact, Parikh
and Wilczek \cite{key-1,key-2} obtained their important result in
the tunnelling framework, which has been an elegant and largely used
approach to obtain Hawking radiation in recent years, see for example
\cite{key-1,key-2,key-19,key-28,key-29,key-30,key-31} and refs. within.
In the tunnelling framework, Hawking's mechanism of particles creation
by BHs \cite{key-3} can be described as tunnelling arising from vacuum
fluctuations near the BH horizon \cite{key-1,key-2,key-19,key-28,key-29,key-30,key-31}.
If a virtual particle pair is created just inside the horizon, the
virtual particle with positive energy can tunnel out. Then, it materializes
outside the BH as a real particle. Analogously, if a virtual particle
pair is created just outside the horizon, the particle with negative
energy can tunnel inwards. In both of the situations, the BH absorbs
the particle with negative energy. The absorptions of such particles
with negative energy are exactly the perturbations generating the
QNMs. The result will be that the BH mass decreases and the particle
with positive energy propagates towards infinity. Thus, subsequent
emissions of quanta appear as Hawking radiation. QNM can be naturally
considered in terms of quantum levels if one interprets the absolute
value of a QNM frequency as the total energy emitted at that level
\cite{key-4,key-5,key-17}. In other words, QNMs frequencies are the
eigenvalues of the system. The Hawking quanta are then interpreted
as the ``jumps'' among the levels. This key point agrees with the
idea that, in an underlying quantum gravity theory, BHs result in
highly excited states. 

We recently used this important issue to re-analyse the spectrum of
BH QNMs through the introduction of a BH \emph{effective temperature
}\cite{key-4,key-5}\emph{.} In our analysis the formula of the horizon's
area quantization and the number of quanta of area resulted to be
functions of the quantum QNMs ``overtone'' number $n$ \cite{key-4,key-5},
i.e. of the BH quantum level. Consequently, Bekenstein-Hawking entropy,
its sub-leading corrections and the number of micro-states resulted
functions of $n\;$ too \cite{key-4,key-5}. 

Here the analysis is improved, removing some approximations that have
been implicitly used in previous works \cite{key-4,key-5} and obtaining
the corrected expressions for the cited formulas like functions of
$n.$ An approximation concerning the maximum value of $n\;$ is also
corrected. As our previous results \cite{key-4,key-5} were strictly
corrected only for scalar and gravitational perturbations, in this
work we show that the analysis holds also for vector perturbations.

We shortly discuss potential important implications also for the BH
information paradox \cite{key-18}.

The analysis in this paper is totally consistent with the general
conviction that BHs result in highly excited states representing both
the ``hydrogen atom'' and the ``quasi-thermal emission'' in quantum
gravity. Previous results in the literature \cite{key-6,key-14,key-15},
obtained in strictly thermal approximation, are consistent with the
results in this paper. 

At the present time, we do not yet have a full theory of quantum gravity.
Thus, we have to be content with the semi-classical approximation.
In fact, as for large $n\:$ \emph{Bohr's correspondence principle}
\cite{key-25,key-26,key-27} holds, such a semi-classical description
is adequate. In this framework, our BH model is somewhat similar to
the semi-classical Bohr's model of the structure of a hydrogen atom
\cite{key-26,key-27}. In our BH model, during a quantum jump a discrete
amount of energy is radiated and for large values of the principal
quantum number $n$ the analysis becomes independent from the other
quantum numbers. In a certain sense, QNMs represent the \textquotedbl{}electron\textquotedbl{}
which jumps from a level to another one and the absolute values of
the QNMs frequencies represent the energy \textquotedbl{}shells\textquotedbl{}.
In Bohr's model \cite{key-26,key-27}, electrons can only gain and
lose energy by jumping from one allowed energy shell to another, absorbing
or emitting radiation with an energy difference of the levels according
to the Planck relation $E=hf$, where $h$ is the Planck constant
and $f$ the transition frequency. In our BH model, QNMs can only
gain and lose energy by jumping from one allowed energy shell to another,
absorbing or emitting radiation (emitted radiation is given by Hawking
quanta) with an energy difference of the levels. On the other hand,
Bohr's model is an approximated model of the hydrogen atom with respect
to the valence shell atom model of full quantum mechanics. In the
same way, our model should be an approximated model of the emitting
BH with respect to the definitive, but at the present time unknown,
model of full quantum gravity theory.

\section{Quasi-normal modes in non-strictly thermal approximation}

\noindent Working with $G=c=k_{B}=\hbar=\frac{1}{4\pi\epsilon_{0}}=1$
(Planck units), in strictly thermal approximation the probability
of emission of Hawking quanta is \cite{key-1,key-2,key-3} 
\begin{equation}
\Gamma\sim\exp(-\frac{\omega}{T_{H}}),\label{eq: hawking probability}
\end{equation}

\noindent where $T_{H}\equiv\frac{1}{8\pi M}$ is the Hawking temperature
and $\omega$ the energy-frequency of the emitted radiation.

\noindent The important correction, due to the BH varying geometry
yields \cite{key-1,key-2}

\noindent 
\begin{equation}
\Gamma\sim\exp[-\frac{\omega}{T_{H}}(1-\frac{\omega}{2M})].\label{eq: Parikh Correction}
\end{equation}

\noindent This result takes into account the BH back reaction and
adds the term $\frac{\omega}{2M}$ like correction \cite{key-1,key-2}.
In a recent paper \cite{key-19} we have improved the tunnelling picture
in \cite{key-1,key-2}. In fact, we have shown that the probability
of emission (\ref{eq: Parikh Correction}) is indeed associated to
the two distributions \cite{key-19} 
\begin{equation}
<n>_{boson}=\frac{1}{\exp\left[-4\pi n\left(M-\omega\right)\omega\right]-1},\;\;<n>_{fermion}=\frac{1}{\exp\left[-4\pi n\left(M-\omega\right)\omega\right]+1},\label{eq: final distributions}
\end{equation}
for bosons and fermions respectively, which are \emph{non} strictly
thermal. 

\noindent As in various fields of physics and astrophysics the deviation
of the spectrum of an emitting body from the strict thermality is
taken into account by introducing an \emph{effective temperature}
(i.e. the temperature of a black body emitting the same total amount
of radiation) we introduced the effective temperature in BH physics
too \cite{key-4,key-5}

\noindent 
\begin{equation}
T_{E}(\omega)\equiv\frac{2M}{2M-\omega}T_{H}=\frac{1}{4\pi(2M-\omega)}.\label{eq: Corda Temperature}
\end{equation}

\noindent One sees that $T_{E}$ depends on the frequency of the emitted
radiation. Therefore, one can rewrite eq. (\ref{eq: Parikh Correction})
in Boltzmann-like form as \cite{key-4,key-5}

\noindent 
\begin{equation}
\Gamma\sim\exp[-\beta_{E}(\omega)\omega]=\exp(-\frac{\omega}{T_{E}(\omega)}),\label{eq: Corda Probability}
\end{equation}

\noindent where $\beta_{E}(\omega)\equiv\frac{1}{T_{E}(\omega)}$
and $\exp[-\beta_{E}(\omega)\omega]$ is the \emph{effective Boltzmann
factor} appropriate for a BH with inverse effective temperature $T_{E}(\omega)$
\cite{key-4,key-5}. The deviation of the BH radiation spectrum from
the strict thermality is given by the ratio $\frac{T_{E}(\omega)}{T_{H}}=\frac{2M}{2M-\omega}$
\cite{key-4,key-5}. The introduction of $T_{E}(\omega)$ permits
the introduction of the \emph{effective mass }and of the \emph{effective
horizon} \cite{key-4,key-5} 
\begin{equation}
M_{E}\equiv M-\frac{\omega}{2},\mbox{ }r_{E}\equiv2M_{E}\label{eq: effective quantities}
\end{equation}

\noindent of the BH \emph{during} the emission of the particle, i.e.
\emph{during} the BH contraction phase \cite{key-4,key-5}. They are
average values of the mass and the horizon \emph{before} and \emph{after}
the emission \cite{key-4,key-5}.

\noindent We have shown that the correction to the thermal spectrum
is also very important for the physical interpretation of BH QNMs,
which, in turn, is very important to realize the underlying quantum
gravity theory as BHs represent theoretical laboratories for developing
quantum gravity and BH QNMs are the best candidates like quantum levels
\cite{key-4,key-5,key-6}. 

\noindent QNMs are radial spin$-j$ perturbations ($j=0,1,2$ for
scalar, vector and gravitational perturbation respectively) of the
Schwarzschild background usually labelled as $\omega_{nl},$ being
$l$ the angular momentum quantum number \cite{key-4,key-5,key-6,key-7}.
For each $l$$\geq2$ for BH perturbations, there is a countable sequence
of QNMs, labelled by the ``overtone'' number $n$ ($n=1,2,...$)
\cite{key-4,key-5,key-7}. For large $n$ the QNMs of the Schwarzschild
BH become independent of $l$. In strictly thermal approximation and
for scalar and gravitational perturbations their frequencies are given
by \cite{key-4,key-5,key-6,key-7}

\noindent 
\begin{equation}
\begin{array}{c}
\omega_{n}=\ln3\times T_{H}+2\pi i(n+\frac{1}{2})\times T_{H}+\mathcal{O}(n^{-\frac{1}{2}})=\\
\\
=\frac{\ln3}{8\pi M}+\frac{2\pi i}{8\pi M}(n+\frac{1}{2})+\mathcal{O}(n^{-\frac{1}{2}}).
\end{array}\label{eq: quasinormal modes}
\end{equation}

\noindent The introduction of the effective temperature $T_{E}(\omega)$
is useful also to analyse BH QNMs \cite{key-4,key-5}. An important
issue is that eq. (\ref{eq: quasinormal modes}) is an approximation
as it assumes that the BH radiation spectrum is strictly thermal.
If one wants to take into account the deviation from the thermal spectrum
the Hawking temperature $T_{H}$ must be replaced by the effective
temperature $T_{E}$ in eq. (\ref{eq: quasinormal modes}) \cite{key-4,key-5}.
Thus, for scalar and gravitational perturbations the correct expression
for the Schwarzschild BH QNMs, which takes into account the non-strict
thermality of the spectrum is \cite{key-4,key-5}

\noindent 
\begin{equation}
\begin{array}{c}
\omega_{n}=\ln3\times T_{E}(\omega_{n})+2\pi i(n+\frac{1}{2})\times T_{E}(\omega_{n})+\mathcal{O}(n^{-\frac{1}{2}})=\\
\\
=\frac{\ln3}{4\pi\left[2M-(\omega_{0})_{n}\right]}+\frac{2\pi i}{4\pi\left[2M-(\omega_{0})_{n}\right]}(n+\frac{1}{2})+\mathcal{O}(n^{-\frac{1}{2}})\simeq\frac{2\pi in}{4\pi\left[2M-(\omega_{0})_{n}\right]},
\end{array}\label{eq: quasinormal modes corrected}
\end{equation}

\noindent where $(\omega_{0})_{n}\equiv|\omega_{n}|.$ We derived
eq. (\ref{eq: quasinormal modes corrected}) in \cite{key-4,key-5}.
A more rigorous derivation of it can be found in detail in the Appendix
of this paper. In that Appendix we also show that the behavior

\noindent 
\begin{equation}
\omega_{n}\simeq\frac{2\pi in}{4\pi\left[2M-(\omega_{0})_{n}\right]}\label{eq: andamento asintotico}
\end{equation}
also holds for $j=1$ (vector perturbations) and this is in full agreement
with Bohr's correspondence principle \cite{key-25,key-26,key-27}
which states that \emph{``transition frequencies at large quantum
numbers should equal classical oscillation frequencies}'' \cite{key-10}. 

\noindent The physical solution for $(\omega_{0})_{n}$ in eq. (\ref{eq: quasinormal modes corrected})
is \cite{key-4,key-5} 
\begin{equation}
(\omega_{0})_{n}=M-\sqrt{M^{2}-\frac{1}{4\pi}\sqrt{(\ln3)^{2}+4\pi^{2}(n+\frac{1}{2})^{2}}}\simeq M-\sqrt{M^{2}-\frac{1}{2}(n+\frac{1}{2})},\label{eq: radice fisica}
\end{equation}
where the term $M-\sqrt{M^{2}-\frac{1}{2}(n+\frac{1}{2})}$ represent
the solution of eq. (\ref{eq: andamento asintotico}), which in turn
hols for $j=0,1,2.$

\section{A note on black hole's remnants}

\noindent As $(\omega_{0})_{n}$ is interpreted like the total energy
emitted at level $n$ \cite{key-4,key-5}, one needs also 

\begin{equation}
M^{2}-\frac{1}{4\pi}\sqrt{(\ln3)^{2}+4\pi^{2}(n+\frac{1}{2})^{2}}\simeq M^{2}-\frac{1}{2}(n+\frac{1}{2})\geq0\label{eq: need}
\end{equation}
in eq. (\ref{eq: radice fisica}). In fact, BHs cannot emit more energy
than their total mass. The expression (\ref{eq: need}) is solved
giving a maximum value for the overtone number $n$ \cite{key-4}

\begin{equation}
n\leq n_{max}=2\pi^{2}\left(\sqrt{16M^{4}-(\frac{\ln3}{\pi})^{2}}-1\right)\simeq2\pi^{2}\pi^{2}(4M^{2}-1),\label{eq: n max}
\end{equation}
which corresponds to $(\omega_{0})_{n_{max}}=M.$ Thus, the countable
sequence of QNMs for emitted energies cannot be infinity although
$n$ can be extremely large \cite{key-4}. On the other hand, we recall
that, by using the Generalized Uncertainty Principle, Adler, Chen
and Santiago \cite{key-8} have shown that the total BH evaporation
is prevented in exactly the same way that the Uncertainty Principle
prevents the hydrogen atom from total collapse. In fact, the collapse
is prevented, not by symmetry, but by dynamics, as the \emph{Planck
distance} and the \emph{Planck mass} are approached \cite{key-8}.
That important result implies that eq. (\ref{eq: need}) has to be
slightly modified, becoming (the \emph{Planck mass} is equal to $1$
in Planck units)

\begin{equation}
M^{2}-\frac{1}{4\pi}\sqrt{(\ln3)^{2}+4\pi^{2}(n+\frac{1}{2})^{2}}\simeq M^{2}-\frac{1}{2}(n+\frac{1}{2})\geq1.\label{eq: need 1}
\end{equation}
By solving eq. (\ref{eq: need 1}) one gets a different value of the
maximum value for the overtone number $n$ 

\begin{equation}
n\leq n_{max}=2\pi^{2}\left(\sqrt{16(M^{2}-1)^{2}-(\frac{\ln3}{\pi})^{2}}-1\right)\simeq2\pi^{2}(4M^{2}-5).\label{eq: n max 1}
\end{equation}
The result in eq. (\ref{eq: n max 1}) improves the one of eq. (\ref{eq: n max})
that we originally derived in \cite{key-4}.

\section{Quasi-normal modes like natural quantum levels}

\noindent Bekenstein \cite{key-9} has shown that the Schwarzschild
BH area quantum is $\triangle A=8\pi$ (the \emph{Planck length} $l_{p}=1.616\times10^{-33}\mbox{ }cm$
is equal to one in Planck units). By analysing Schwarzschild BH QNs,
Hod found a different numerical coefficient \cite{key-10}. Hod's
work was refined by Maggiore \cite{key-6}, who re-obtained the original
result by Bekenstein. In \cite{key-4,key-5} we further improved the
result by Maggiore taking into account the deviation from the strictly
thermal feature. In fact, in \cite{key-4,key-5} we used eq. (\ref{eq: quasinormal modes corrected})
instead of eq. (\ref{eq: quasinormal modes}) \cite{key-4,key-5}.
From eq. (\ref{eq: radice fisica}) one gets that an emission involving
$n$ and $n-1$  gives a variation of energy \cite{key-4,key-5} 
\begin{equation}
\triangle M_{n}=(\omega_{0})_{n-1}-(\omega_{0})_{n}=-f_{n}(M,n)\label{eq: variazione}
\end{equation}

\noindent where one defines \cite{key-4,key-5}

\noindent 
\begin{equation}
\begin{array}{c}
f_{n}(M,n)\equiv\\
\\
\sqrt{M^{2}-\frac{1}{4\pi}\sqrt{(\ln3)^{2}+4\pi^{2}(n-\frac{1}{2})^{2}}}-\sqrt{M^{2}-\frac{1}{4\pi}\sqrt{(\ln3)^{2}+4\pi^{2}(n+\frac{1}{2})^{2}}}\simeq\\
\\
\simeq\sqrt{M^{2}-\frac{1}{2}(n-\frac{1}{2})}-\sqrt{M^{2}-\frac{1}{2}(n+\frac{1}{2})}
\end{array}\label{eq: f(M,n)}
\end{equation}

\noindent Together with collaborators, we have recently shown that
the results (\ref{eq: variazione}) (\ref{eq: f(M,n)}) also hold
for Kerr BHs in the case $M^{2}\gg J,$ where $J$ is the angular
momentum of the BH \cite{key-17}.

\noindent Recalling that, as for the Schwarzschild BH the \emph{horizon
area} $A$ is related to the mass through the relation $A=16\pi M^{2},$
a variation $\triangle M\,$ in the mass generates a variation

\noindent 
\begin{equation}
\triangle A=32\pi M\triangle M\label{eq: variazione area}
\end{equation}

\noindent in the area. Combining eqs. (\ref{eq: variazione area})
and (\ref{eq: variazione}) one gets \cite{key-4,key-5}

\noindent 
\begin{equation}
\triangle A=32\pi M\triangle M_{n}=-32\pi M\times f_{n}(M,n).\label{eq: area quantum}
\end{equation}

\noindent As we consider large $n,\:$ eq. (\ref{eq: f(M,n)}) is
well approximated by $f_{n}(M,n)\approx\frac{1}{4M}$ \cite{key-4,key-5}
and eq. (\ref{eq: area quantum}) becomes $\triangle A\approx-8\pi$
which is the original result by Bekenstein for the area quantization
(a part a sign because we consider an emission instead of an absorption).
Then, only when $n$ is enough large the levels are approximately
equally spaced \cite{key-4,key-5}. Instead, for smaller $n\,$ there
are deviations.

\noindent One assumes that, for large $n$, the horizon area is quantized
\cite{key-4,key-5,key-6} with a quantum $|\triangle A|.$ The total
horizon area must be $A=N|\triangle A|$ where the integer $N$ is
the number of quanta of area. One gets \cite{key-4,key-5}

\noindent 
\begin{equation}
N=\frac{A}{|\triangle A|}=\frac{16\pi M^{2}}{32\pi M\cdot f_{n}(M,n)}=\frac{M}{2f_{n}(M,n)}.\label{eq: N}
\end{equation}
This permits to write the famous formula of Bekenstein-Hawking entropy
\cite{key-3,key-9} like a function of the quantum overtone number
$n$ \cite{key-4,key-5}

\noindent 
\begin{equation}
S_{BH}=\frac{A}{4}=8\pi NM\cdot f_{n}(M,n).\label{eq: Bekenstein-Hawking}
\end{equation}

\noindent As we consider large $n,$ the approximation $f_{n}(M,n)\approx\frac{1}{4M}\:$
permits to re-obtain the standard result \cite{key-6,key-11,key-12,key-13}

\noindent 
\begin{equation}
S_{BH}\rightarrow2\pi N,\label{eq: entropia circa}
\end{equation}

\noindent like good approximation. Recent results show that that BH
entropy contains three parts which are important to realize the underlying
quantum gravity theory. They are the Bekenstein-Hawking entropy and
two sub-leading corrections: the logarithmic term and the inverse
area term \cite{key-14,key-15}

\noindent 
\begin{equation}
S_{total}=S_{BH}-\ln S_{BH}+\frac{3}{2A}.\label{eq: entropia totale}
\end{equation}

\noindent Thus, eq. (\ref{eq: Bekenstein-Hawking}) permits to re-write
eq. (\ref{eq: entropia totale}) like \cite{key-4,key-5}

\noindent 
\begin{equation}
S_{total}=8\pi NM\cdot f_{n}(M,n)-\ln\left[8\pi NM\cdot f_{n}(M,n)\right]+\frac{3}{64\pi NM\cdot f_{n}(M,n)}\label{eq: entropia totale 2}
\end{equation}

\noindent that, as one considers large $n,\:$ is well approximated
by \cite{key-4,key-5}

\noindent 
\begin{equation}
S_{total}\simeq2\pi N-\ln2\pi N+\frac{3}{16\pi N}.\label{eq: entropia totale approssimata}
\end{equation}

\noindent Thus, at a level $n-1,$ the BH has a number of micro-states
\cite{key-4,key-5}

\noindent 
\begin{equation}
g(N)\propto\exp\left\{ 8\pi NM\cdot f_{n}(M,n)-\ln\left[8\pi NM\cdot f_{n}(M,n)\right]+\frac{3}{64\pi NM\cdot f_{n}(M,n)}\right\} ,\label{eq: microstati}
\end{equation}

\noindent that, for large $n,\:$ is well approximated by \cite{key-4,key-5}

\noindent 
\begin{equation}
g(N)\propto\exp\left[2\pi N-\ln\left(2\pi N\right)+\frac{3}{16\pi N}\right].\label{eq: microstati circa}
\end{equation}
Eqs. (\ref{eq: entropia circa}), (\ref{eq: entropia totale approssimata})
and (\ref{eq: microstati circa}) are in agreement with previous literature
\cite{key-6,key-14,key-15}, in which the strictly thermal approximation
were used. 

Actually, in previous discussion, and hence also in \cite{key-4,key-5},
we used an implicit simplification. Now, we improve the analysis by
removing such a simplification and by giving the correct results.

In fact, we note that, after an high number of emissions (and potential
absorptions as the BH can capture neighboring particles), the BH mass
changes from $M\;$ to 

\begin{equation}
M_{n-1}\equiv M-(\omega_{0})_{n-1},\label{eq: me-1}
\end{equation}
where $(\omega_{0})_{n-1}$ is the total energy emitted by the BH
at that time, and the BH is excited at a level $n-1$. In the transition
from the state with $n-1$ to the state with $n\;$ the BH mass changes
again from $M_{n-1}$ to

\begin{equation}
M_{n}\equiv M-(\omega_{0})_{n-1}+\triangle M_{n},\label{eq: me}
\end{equation}
which, by using eq. (\ref{eq: variazione}), becomes 
\begin{equation}
\begin{array}{c}
M_{n}=M-(\omega_{0})_{n-1}-f_{n}(M,n)=\\
\\
=M-(\omega_{0})_{n-1}+(\omega_{0})_{n-1}-(\omega_{0})_{n}=M-(\omega_{0})_{n}.
\end{array}\label{eq: me 2}
\end{equation}
Now, the BH is excited at a level $n$. By considering eq. (\ref{eq: radice fisica}),
eqs. (\ref{eq: me-1}) and (\ref{eq: me 2}) read

\begin{equation}
M_{n-1}=\sqrt{M^{2}-\frac{1}{4\pi}\sqrt{(\ln3)^{2}+4\pi^{2}(n-\frac{1}{2})^{2}}}\simeq\sqrt{M^{2}-\frac{1}{2}(n-\frac{1}{2})}\label{eq: me 3}
\end{equation}
and 

\begin{equation}
M_{n}=\sqrt{M^{2}-\frac{1}{4\pi}\sqrt{(\ln3)^{2}+4\pi^{2}(n+\frac{1}{2})^{2}}}\simeq\sqrt{M^{2}-\frac{1}{2}(n+\frac{1}{2})}.\label{eq: me 4}
\end{equation}
For extremely large $n\quad$ the condition $M_{n}\simeq M$, that
we implicitly used in previous discussion, does not hold because the
BH has emitted a large amount of mass. This implies that, if one uses
eqs. (\ref{eq: radice fisica}) and eq. (\ref{eq: me 3}), eq. (\ref{eq: area quantum})
has to be correctly rewritten as

\begin{equation}
\triangle A_{n-1}\equiv32\pi M_{n-1}\triangle M_{n}=-32\pi M_{n-1}\times f_{n}(M,n)\label{eq: area quantum e}
\end{equation}
This equation should give the area quantum of an excited BH for an
emission from the level $n-1$ to the level $n$ in function of the
quantum number $n$ and of the initial BH mass. Actually, there is
a problem in eq. (\ref{eq: area quantum e}). In fact, an absorption
from the level $n$ to the level $n-1$ is now possible, with an absorbed
energy \cite{key-5} 
\begin{equation}
(\omega_{0})_{n}-(\omega_{0})_{n-1}=f_{n}(M,n)=-\triangle M_{n}.\label{eq: absorbed}
\end{equation}
In that case, the quantum of area should be 
\begin{equation}
\triangle A_{n}\equiv-32\pi M_{n}\triangle M_{n}=32\pi M_{n}\times f_{n}(M,n),\label{eq: area quantum a}
\end{equation}
and the absolute value of the area quantum for an absorption from
the level $n$ to the level $n-1$ is different from the absolute
value of the area quantum for an emission from the level $n-1$ to
the level $n\;$ because $M_{n-1}\neq M_{n}.$ Clearly, one indeed
expects the area spectrum to be the same for absorption and emission.
This inconsistency is solved if, once again, one considers the \emph{effective
mass} which correspond to the transitions between the two levels $n\;$
and $n-1$, which is the same for emission and absorption

\begin{equation}
\begin{array}{c}
M_{E(n,\; n-1)}\equiv\frac{1}{2}\left(M_{n-1}+M_{n}\right)=\\
\\
=\frac{1}{2}\left(\sqrt{M^{2}-\frac{1}{4\pi}\sqrt{(\ln3)^{2}+4\pi^{2}(n-\frac{1}{2})^{2}}}+\sqrt{M^{2}-\frac{1}{4\pi}\sqrt{(\ln3)^{2}+4\pi^{2}(n+\frac{1}{2})^{2}}}\right)\simeq\\
\\
\simeq\frac{1}{2}\left(\sqrt{M^{2}-\frac{1}{2}(n-\frac{1}{2})}+\sqrt{M^{2}+\frac{1}{2}(n-\frac{1}{2})}\right)
\end{array}.\label{eq: massa effettiva n}
\end{equation}
Replacing $M_{n-1}$ with $M_{E(n,\; n-1)}$ in eq. (\ref{eq: area quantum e})
and $M_{n}$ with $M_{E(n,\; n-1)}$ in eq. (\ref{eq: area quantum a})
we obtain 
\begin{equation}
\begin{array}{c}
\triangle A_{n-1}\equiv32\pi M_{E(n,\; n-1)}\triangle M_{n}=-32\pi M_{E(n,\; n-1)}\times f_{n}(M,n)\qquad emission\\
\\
\triangle A_{n}\equiv-32\pi M_{E(n,\; n-1)}\triangle M_{n}=32\pi M_{E(n,\; n-1)}\times f_{n}(M,n)\qquad absorption
\end{array}\label{eq: expects}
\end{equation}
and now one gets $\alpha=|\triangle A_{n}|=|\triangle A_{n-1}|.$
By using eqs. (\ref{eq: f(M,n)}) and (\ref{eq: massa effettiva n})
one finds 

\begin{equation}
\begin{array}{c}
\alpha=|\triangle A_{n}|=|\triangle A_{n-1}|\\
\\
=4\left(\sqrt{(\ln3)^{2}+4\pi^{2}(n+\frac{1}{2})^{2}}-\sqrt{(\ln3)^{2}+4\pi^{2}(n-\frac{1}{2})^{2}}\right)\simeq8\pi.
\end{array}\label{eq: 8 pi planck}
\end{equation}
Hence, the introduction of the effective temperature and of the effective
mass does not degrade the importance of the Hawking temperature. Indeed,
the effective temperature and the effective mass are introduced because
the values of the Hawking temperature and of the mass change with
discrete behavior in time. Thus, it is not clear which value of the
Hawking temperature and of the mass have to be associated to the emission
(or to the absorption) of the particle. Has one to consider the values
of the Hawking temperature and of the mass \emph{before} the emission
(absorption) or the values \emph{after} the emission (absorption)?
The answer is that one must consider \emph{intermediate} values, the
effective temperature and the effective mass. In a certain sense,
they represent the values of the BH temperature and BH mass\emph{
during} the emission. 

We note that, as we consider large $n$, the result (\ref{eq: 8 pi planck})
is $\simeq8\pi,$ which also holds for vector perturbations. Thus,
one can take the result (\ref{eq: 8 pi planck}) as the quantization
of the area of the horizon of a Schwarzschild BH. Again, the original
famous result by Bekenstein \cite{key-9} is a good approximation
and a confirmation of the correctness of the current analysis. This
is also in full agreement with Bohr's correspondence principle \cite{key-25,key-26,key-27}.

In previous analysis and in \cite{key-4,key-5}, the simplification
$(\omega_{0})_{n-1}\ll M$ has been implicitly used, i.e. the energy
associated to the QNM is much less than the original mass-energy of
the BH. Clearly, in that case the correction given by eq. (\ref{eq: area quantum e})
results non-essential, as one can neglect the difference between the
initial BH mass $M$ and the mass of the excited BH $M_{n-1}$, but
it becomes very important when $M_{n}\simeq M\;$ does not hold, i.e.
for very highly excited BHs. In that case, for example in the latest
stages of the BH evaporation (but before arriving at the Planck scale,
where our semi-classical approximation breaks down and a full quantum
gravity theory is needed), it could be $(\omega_{0})_{n}\lesssim M$,
and further corrections on previous formulas are needed. Putting $A_{n-1}\equiv16\pi M_{n-1}^{2}$
and $A_{n}\equiv16\pi M_{n}^{2}$, the formulas of the number of quanta
of area and of the Bekenstein-Hawking entropy become 

\noindent 
\begin{equation}
N_{n-1}\equiv\frac{A_{n-1}}{|\triangle A_{n-1}|}=\frac{16\pi M_{n-1}^{2}}{32\pi M_{E(n,\; n-1)}\cdot f_{n}(M,n)}=\frac{M_{n-1}^{2}}{2M_{E(n,\; n-1)}\cdot f_{n}(M,n)}\label{eq: N n-1}
\end{equation}

\noindent before the emission, and
\begin{equation}
N_{n}\equiv\frac{A_{n}}{|\triangle A_{n}|}=\frac{16\pi M_{n}^{2}}{M_{E(n,\; n-1)}}=\frac{M_{n}^{2}}{2M_{E(n,\; n-1)}\cdot f_{n}(M,n)}\label{eq: N n}
\end{equation}
after the emission respectively. One can easily check that

\noindent 
\begin{equation}
N_{n}-N_{n-1}=\frac{M_{n}^{2}-M_{n-1}^{2}}{2M_{E(n,\; n-1)}\cdot f_{n}(M,n)}=\frac{f_{n}(M,n)\left(M_{n-1}+M_{n}\right)}{2M_{E(n,\; n-1)}\cdot f_{n}(M,n)}=1\label{eq: check}
\end{equation}
as one expects. Hence, the formulas of the Bekenstein-Hawking entropy
read 
\begin{equation}
\begin{array}{c}
\left(S_{BH}\right)_{n-1}\equiv\frac{A_{n-1}}{4}=8\pi N_{n-1}M_{n-1}|\triangle M_{n}|=8\pi N_{n-1}M_{n-1}\cdot f_{n}(M,n)\\
\\
=4\pi\left(M^{2}-\frac{1}{4\pi}\sqrt{(\ln3)^{2}+4\pi^{2}(n-\frac{1}{2})^{2}}\right),
\end{array}\label{eq: Bekenstein-Hawking  n-1}
\end{equation}
before the emission and 
\begin{equation}
\begin{array}{c}
\left(S_{BH}\right)_{n}\equiv\frac{A_{n}}{4}=8\pi N_{n}M_{n}|\triangle M_{n}|=8\pi N_{n}M_{n}\cdot f_{n}(M,n)\\
\\
=4\pi\left(M^{2}-\frac{1}{4\pi}\sqrt{(\ln3)^{2}+4\pi^{2}(n+\frac{1}{2})^{2}}\right)
\end{array}\label{eq: Bekenstein-Hawking  n}
\end{equation}
after the emission respectively. Notice that, as $n\gg1$, one obtains
\begin{equation}
\left(S_{BH}\right)_{n}\simeq\left(S_{BH}\right)_{n-1}\simeq4\pi\left(M^{2}-\frac{n}{2}\right).\label{eq: sb circa}
\end{equation}

\noindent Again, formula (\ref{eq: sb circa}) works for all $j=0,1,2.$
The formulas of the total entropy that takes into account the sub-leading
corrections to Bekenstein-Hawking entropy become

\noindent 
\begin{equation}
\begin{array}{c}
\left(S_{total}\right)_{n-1}=8\pi N_{n-1}M_{n-1}\cdot f_{n}(M,n)\\
\\
-\ln\left[8\pi N_{n-1}M_{n-1}\cdot f_{n}(M,n)\right]+\frac{3}{64\pi N_{n-1}M_{n-1}\cdot f_{n}(M,n)}
\end{array}\label{eq: entropia n-1}
\end{equation}

\noindent before the emission, and 
\begin{equation}
\begin{array}{c}
\left(S_{total}\right)_{n}=8\pi N_{n}M_{n}\cdot f_{n}(M,n)\\
\\
-\ln\left[8\pi N_{n}M_{n}\cdot f_{n}(M,n)\right]+\frac{3}{64\pi N_{n}M_{n}\cdot f_{n}(M,n)}
\end{array}\label{eq: entropia n}
\end{equation}
after the emission, respectively.

Hence, at level $n-1$ the BH has a number of micro-states 
\begin{equation}
\begin{array}{c}
g(N_{n-1})\propto\exp\{8\pi N_{n-1}M_{n-1}\cdot f_{n}(M,n)+\\
\\
-\ln\left[8\pi N_{n-1}M_{n-1}\cdot f_{n}(M,n)\right]+\frac{3}{64\pi N_{n-1}M_{n-1}\cdot f_{n}(M,n)}
\end{array}\label{eq: microstati n-1}
\end{equation}
and, at level $n$, after the emission, the number of micro-states
is 
\begin{equation}
\begin{array}{c}
g(N_{n})\propto\exp\{8\pi N_{n}M_{n}\cdot f_{n}(M,n)+\\
\\
-\ln\left[8\pi N_{n}M_{n}\cdot f_{n}(M,n)\right]+\frac{3}{64\pi N_{n}M_{n}\cdot f_{n}(M,n)}
\end{array}\label{eq: microstati n}
\end{equation}
All these corrections, which represent the correct formulas of an
excited BH for a transition between the levels $n-1$ and $n$ in
function of the quantum number $n$, result very important for very
highly excited BHs, when $n$ becomes extremely large and $M_{n}\simeq M\;$
does not hold (in particular in the last stages of the BH evaporation,
before arriving at the Planck scale, when $(\omega_{0})_{n}\lesssim M$).

Instead, when $(\omega_{0})_{n}\ll M$, formulas (\ref{eq: area quantum}),
(\ref{eq: N}), (\ref{eq: entropia totale 2}) and (\ref{eq: microstati}),
are a good approximation. Such formulas are a better approximation
with respect to formulas (\ref{eq: entropia circa}), (\ref{eq: entropia totale approssimata})
and (\ref{eq: microstati circa}) which were used in previous results
in the literature \cite{key-6,key-14,key-15}. In those works the
strictly thermal approximation was used. We also see that, for $(\omega_{0})_{n-1}\ll M$,
$M_{n}\simeq M_{n-1}\simeq M$, and eqs. (\ref{eq: area quantum}),
(\ref{eq: N}), (\ref{eq: entropia totale 2}) and (\ref{eq: microstati}),
are easily recovered.

\section{Final discussion and conclusion remarks}

We explain the way in which our BH model works. Let us consider a
BH original mass $M.$ After an high number of emissions (and potential
absorptions as the BH can capture neighboring particles), the BH will
be at an excited level $n-1$ and its mass will be $M_{n-1}\equiv M-(\omega_{0})_{n-1}$
where $(\omega_{0})_{n-1},$ is the absolute value of the frequency
of the QNM associated to the excited level $n-1.\;$ $(\omega_{0})_{n-1}\;$
is also the total energy emitted at that time. The BH can further
emit an energy to jump to the subsequent level: $\triangle M_{n}=(\omega_{0})_{n-1}-(\omega_{0})_{n}.$
Now, the BHat an excited level $n$ and the BH mass will be 
\begin{equation}
\begin{array}{c}
M_{n}\equiv M-(\omega_{0})_{n-1}+\triangle M_{n-1}=\\
\\
=M-(\omega_{0})_{n-1}+(\omega_{0})_{n-1}-(\omega_{0})_{n}=M-(\omega_{0})_{n}.
\end{array}\label{eq: masse}
\end{equation}
The BH can, in principle, return to the level $n-1$ by absorbing
an energy $-\triangle M_{n}=(\omega_{0})_{n}-(\omega_{0})_{n-1}$.
We have also shown that the quantum of area is \emph{the same} for
both absorption and emission, given by eq. (\ref{eq: 8 pi planck}),
as one expects. 

There are three different physical situations for excited BHs ($n\gg1$):
\begin{enumerate}
\item $n\;$ is large, but not enough large. It is also $(\omega_{0})_{n}\ll M_{n}\simeq M$
and one can use eqs. (\ref{eq: area quantum}), (\ref{eq: Bekenstein-Hawking}),
(\ref{eq: entropia totale 2}) and (\ref{eq: microstati}) which results
a better approximation than eqs. (\ref{eq: entropia circa}), (\ref{eq: entropia totale approssimata})
and (\ref{eq: microstati circa}) which were used in previous literature
in strictly thermal approximation \cite{key-6,key-14,key-15}
\item $n\;$ is very much larger than in point 1, but before arriving at
the Planck scale. In that case, it can be $(\omega_{0})_{n}\lesssim M,$
while $M_{n}\simeq M$ does not hold and one must use the eqs. (37),
and from (41) to (47).
\item At the Planck scale $n\;$ is larger also than in point 2, we need
a full theory of quantum gravity.
\end{enumerate}
In summary, in this paper we analyzed BH QNMs in terms of quantum
levels following the idea that, in an underlying quantum gravity theory,
BHs result in highly excited states. By using the concept of effective
temperature, we took into account the important issue that QNMs spectrum
is not strictly thermal.

The obtained results improve our previous analysis in \cite{key-4,key-5}
because here we removed some approximations that we implicitly used
in \cite{key-4,key-5}. The results look particularly intriguing as
important modifies on BH quantum physics have been realized. In fact,
we found the correct formulas of the horizon's area quantization and
of the number of quanta of area like functions of the quantum ``overtone''
number $n$. Consequently, we also found the correct expressions of
Bekenstein-Hawking entropy, its sub-leading corrections and the number
of micro-states, i.e. quantities which are fundamental to realize
the underlying quantum gravity theory, like functions of the quantum
``overtone'' number $n$. In other words, the cited important quantities
result to depend on the excited BH quantum state. An approximation
concerning the maximum value of $n\;$ has been also corrected. As
our previous results in \cite{key-4,key-5} were strictly corrected
only for scalar and gravitational perturbations, the results of this
work have shown that the analysis holds also for vector perturbations.

We stress that the analysis is totally consistent with the general
conviction that BHs result in highly excited states representing both
the ``hydrogen atom'' and the ``quasi-thermal emission'' in quantum
gravity and that previous results in the literature, where the thermal
approximation has been used \cite{key-6,key-14,key-15}, are consistent
with the results in this paper. As at the present time we do not yet
have a full theory of quantum gravity. Then, we must be content with
the semi-classical approximation. As for large $n$ Bohr's correspondence
principle holds \cite{key-25,key-26,key-27}, such a semi-classical
description is adequate. Thus, we can consider an intriguing analogy
in which our BH model is somewhat similar to the semi-classical Bohr's
model of the structure of a hydrogen atom \cite{key-26,key-27}. In
our BH model, during a quantum jump a discrete amount of energy is
indeed radiated and for large values of the principal quantum number
$n$ the analysis becomes independent from the other quantum numbers.
In a certain sense, QNMs represent the \textquotedbl{}electron\textquotedbl{}
which jumps from a level to another one and the absolute values of
the QNMs frequencies represent the energy \textquotedbl{}shells\textquotedbl{}.
In Bohr's model \cite{key-26,key-27}, electrons can only gain and
lose energy by jumping from one allowed energy shell to another, absorbing
or emitting radiation with an energy difference of the levels according
to the Planck relation $E=hf$, where $\: h\:$ is the Planck constant
and $f\:$ the transition frequency. In our BH model, QNMs can only
gain and lose energy by jumping from one allowed energy shell to another,
absorbing or emitting radiation (emitted radiation is given by Hawking
quanta) with an energy difference of the levels according to eqs.
(\ref{eq: variazione}) and (\ref{eq: f(M,n)}). On the other hand,
Bohr's model is an approximated model of the hydrogen atom with respect
to the valence shell atom model of full quantum mechanics. In the
same way, our BH model should be an approximated model of the emitting
BH with respect to the definitive, but at the present time unknown,
model of full quantum gravity theory. 

It is also important to emphasize that the results in this paper could
have important implications for the BH information paradox. The more
important arguments that information can be lost during BH evaporation
rely indeed on the assumption of strict thermal behavior of the spectrum
\cite{key-18}. On the other hand, the results in this paper show
that BHs seem to be well defined quantum mechanical systems, which
have ordered, discrete quantum spectra. This important point is surely
consistent with the unitarity of the underlying quantum gravity theory
and endorses the idea that information should come out in BH evaporation
\cite{key-20,key-21,key-22,key-23,key-24,key-32,key-33}.

\section{Acknowledgements}

It is a pleasure to thank my colleague and collaborator Hossein Hendi
and my students Reza Katebi and Nathan Schmidt for useful discussions
on BH physics.

\section*{Appendix. Derivation of the quasi-normal modes equation in non-strictly
thermal approximation}

\noindent Being frequencies of the radial spin$-j=0,1,2$ perturbations
$\phi$ of the Schwarzschild space-time, QNMs are governed by the
Schrodinger-like equation \cite{key-4,key-5,key-16} 
\begin{equation}
\left(-\frac{\partial^{2}}{\partial x^{2}}+V(x)-\omega^{2}\right)\phi.\label{eq: diff.}
\end{equation}
In strictly thermal approximation one introduces the Regge-Wheeler
potential

\noindent 
\begin{equation}
V(x)=V\left[x(r)\right]=\left(1-\frac{2M}{r}\right)\left(\frac{l(l+1)}{r^{2}}+2\frac{(1-j^{2})M}{r^{3}}\right).\label{eq: Regge-Wheeler-1}
\end{equation}
We recall that $j=0,1,2$ for scalar, vector and gravitational perturbation
respectively.

\noindent The relation between the Regge-Wheeler ``tortoise'' coordinate
$x$ and the radial coordinate $r\;$ is \cite{key-4,key-5,key-16}

\noindent 
\begin{equation}
\begin{array}{c}
x=r+2M\ln\left(\frac{r}{2M}-1\right)\\
\\
\frac{\partial}{\partial x}=\left(1-\frac{2M}{r}\right)\frac{\partial}{\partial r}.
\end{array}\label{eq:original  tortoise}
\end{equation}

\noindent These states are analogous to quasi-stationary states in
quantum mechanics \cite{key-16}. Thus, their frequency is allowed
to be complex \cite{key-16}. They must have purely outgoing boundary
conditions both at the horizon ($r=2M$) and in the asymptotic region
($r\rightarrow\infty$) \cite{key-16} 

\noindent 
\begin{equation}
\phi(x)\sim c_{\pm}\exp\left(\mp i\omega x\right)\quad for\quad x=\pm\infty.\label{eq: condizioni contorno}
\end{equation}

\noindent Considering the non-strictly thermal behavior of BHs, one
substitutes the original BHs $M$ in eqs. (\ref{eq: diff.}) and (\ref{eq: Regge-Wheeler-1})
with the effective mass of the contracting BH defined in eq. (\ref{eq: effective quantities})
\cite{key-4,key-5}. Hence, eqs. (\ref{eq: Regge-Wheeler-1}) and
(\ref{eq:original  tortoise}) are replaced by the \emph{effective
equations }\cite{key-4,key-5} 
\begin{equation}
V(x)=V\left[x(r)\right]=\left(1-\frac{2M_{E}}{r}\right)\left(\frac{l(l+1)}{r^{2}}+2\frac{(1-j^{2})M}{r^{3}}\right)\label{eq: effettiva 1}
\end{equation}

\noindent and 
\begin{equation}
\begin{array}{c}
x=r+2M_{E}\ln\left(\frac{r}{2M_{E}}-1\right)\\
\\
\frac{\partial}{\partial x}=\left(1-\frac{2M_{E}}{r}\right)\frac{\partial}{\partial r}.
\end{array}\label{eq: effettiva 2}
\end{equation}

\noindent In order to streamline the formulas, here we also set 
\begin{equation}
2M_{E}=r_{E}\equiv1\;\;\;\; and\;\;\;\; m\equiv n+1.\label{eq: set}
\end{equation}

\noindent As the \emph{Planck mass} $m_{p}\:$ is equal to $1$ in
Planck units one rewrites (\ref{eq: quasinormal modes corrected})
as 
\begin{equation}
\frac{\omega_{m}}{m_{p}^{2}}=\frac{\ln3}{4\pi}+\frac{i}{2}(m-\frac{1}{2})+\mathcal{O}(m^{-\frac{1}{2}}),\quad for\: m\gg1,\label{eq: intuizione}
\end{equation}
where now $m_{p}\neq1.$ Putting

\noindent 
\begin{equation}
\tilde{\omega}_{m}\equiv\frac{\omega_{m}}{m_{p}^{2}},\label{eq: important definition}
\end{equation}

\noindent eqs. (\ref{eq: quasinormal modes corrected}), (\ref{eq: diff.}),
(\ref{eq: effettiva 1}) and (\ref{eq: effettiva 2}) become 
\begin{equation}
\tilde{\omega}_{m}=\frac{\ln3}{4\pi}+\frac{i}{2}(m-\frac{1}{2})+\mathcal{O}(m^{-\frac{1}{2}}),\quad for\: m\gg1,\label{eq: rottura}
\end{equation}

\noindent 
\begin{equation}
\left(-\frac{\partial^{2}}{\partial x^{2}}+V(x)-\tilde{\omega}^{2}\right)\phi,\label{eq: rottura 0}
\end{equation}

\noindent 
\begin{equation}
V(x)=V\left[x(r)\right]=\left(1-\frac{1}{r}\right)\left(\frac{l(l+1)}{r^{2}}-\frac{3(1-j^{2})}{r^{3}}\right)\label{eq: rottura 2}
\end{equation}

\noindent and 
\begin{equation}
\begin{array}{c}
x=r+\ln\left(r-1\right)\\
\\
\frac{\partial}{\partial x}=\left(1-\frac{1}{r}\right)\frac{\partial}{\partial r}
\end{array}\label{eq: rottura 3}
\end{equation}
respectively. Although $M_{E}$ and $r_{E}$ (and consequently the
tortoise coordinate and the Regge-Wheeler potential) are frequency
dependent, eq. (\ref{eq: set}) translates such a frequency dependence
into a continually rescaled mass unit in the following discussion.
We will show at the end of this Appendix that such a rescaling is
extremely slow and always included within a factor $2.$ Thus, it
does not influence the following analysis. 

\noindent We emphasize that here after we closely follow ref. \cite{key-16}.
The solution of eq. (\ref{eq: rottura 0}) can be expanded as \cite{key-16}

\begin{equation}
\phi(r)=\left(\frac{r-1}{r^{2}}\right)^{i\tilde{\omega}}\exp\left[\text{-}i\text{\ensuremath{\tilde{\omega}}}(r\text{-}1)\right]\sum_{m}a_{m}\left(\frac{r-1}{r}\right)^{m}.\label{eq: fi di r}
\end{equation}
The pre-factor has to satisfy the boundary conditions (\ref{eq: condizioni contorno})
both at the effective horizon ($r=1$) and in the asymptotic region
($r=\infty$) \cite{key-16}:
\begin{enumerate}
\item One needs $\exp\left[-i\text{\ensuremath{\tilde{\omega}}}(r-1)\right]$
for the correct leading evolution at $r\rightarrow\infty$
\item $\left(r-1\right)^{i\tilde{\omega}}$ fixes the evolution at $r\rightarrow1^{+}$
\item $\left(\frac{1}{r^{2}}\right)^{i\tilde{\omega}}$ arranges the sub-leading
evolution at $r\rightarrow\infty$ which arises from the logarithmic
term in eq. (\ref{eq: rottura 3}).
\end{enumerate}
The power series (\ref{eq: fi di r}) converges for $\frac{1}{2}<r\leq\infty$
assuming that the boundary conditions at $r=\infty$ are preserved
\cite{key-16}. On the other hand, eq. (\ref{eq: diff.}) is equivalent
to the recursion relation \cite{key-16}

\begin{equation}
c_{0}(m,\tilde{\omega})a_{m}+c_{1}(m,\text{\ensuremath{\tilde{\omega}}})a_{m}-1+c_{2}(m,\text{\ensuremath{\tilde{\omega}}})a_{m}-2=0.\label{eq: recursione}
\end{equation}
One can extract the coefficients $\; c_{k}(m,\tilde{\omega})$ from
eq. (\ref{eq: recursione}) and rewrite them in a more convenient
way \cite{key-16}: 

\begin{equation}
c_{0}(m,\text{\ensuremath{\tilde{\omega}}})=m(m+2i\text{\ensuremath{\tilde{\omega}}})\label{eq: c0}
\end{equation}

\begin{equation}
c_{1}(m,\tilde{\omega})=-2(m+2i\text{\ensuremath{\tilde{\omega}}}-\frac{1}{2})^{2}+j^{2}-\frac{1}{2}\label{eq:c1}
\end{equation}

\begin{equation}
c_{2}(m,\tilde{\omega})=(m+2i\text{\ensuremath{\tilde{\omega}}}-1)^{2}-j^{2}.\label{eq: c3}
\end{equation}
We see that, except for $c_{0},$ the coefficients $c_{k}$ depend
on $m,\tilde{\omega}\:$ only through their combination $m+2i\text{\ensuremath{\tilde{\omega}}}.$
The preservation of the boundary conditions at $r=1\;$ is guaranteed
by the initial conditions for the recursion relation \cite{key-16}.
Such conditions are $a_{0}=1$ (in general any non-zero constant that
we set equal to the unity for the sake of simplicity), and $a_{-1}=0$
(that, together with eq. (\ref{eq: recursione}) implies $a_{\lyxmathsym{\textminus}m}=0$
for all positive $m$) \cite{key-15}. One also defines \cite{key-16}

\begin{equation}
R_{m}=-\frac{a_{m}}{a_{m-1}},\label{eq: Rm}
\end{equation}
where we choose the minus sign in agreement with \cite{key-16}. By
using eq. (\ref{eq: recursione}), one gets \cite{key-16}

\begin{equation}
c_{1}(m,\tilde{\omega})-c_{0}(m,\text{\ensuremath{\tilde{\omega}}})R_{m}=\frac{c_{2}(m,\tilde{\omega})}{R_{m-1}}\label{eq: 1-0}
\end{equation}
which can be rewritten as

\begin{equation}
R_{m-1}=\frac{c_{2}(m,\tilde{\omega})}{c_{1}(m,\tilde{\omega})-c_{0}(m,\text{\ensuremath{\tilde{\omega}}})R_{m}}.\label{eq: 1-0 bis}
\end{equation}
Therefore, one can write $R_{m}\:$ in terms of a continued fraction.
The condition $a_{-1}=0\:$ becomes \cite{key-16}

\begin{equation}
R_{0}=\infty\;\;\;\rightarrow\;\;\; c_{1}(1,\tilde{\omega})-c_{0}(1,\text{\ensuremath{\tilde{\omega}}})R_{1}=0.\label{eq: Therefore}
\end{equation}
As eq. (\ref{eq: fi di r}) converges at $r=\infty,$ there is a particular
asymptotic form of $R_{m}\:$ for large $m$ (with $|R_{m}|<1$ for
very large $m$) \cite{key-16}. Thus, the boundary conditions require
eq. (\ref{eq: Therefore}). One can write down eq. (\ref{eq: Therefore})
in terms of continued fractions \cite{key-16}

\begin{equation}
0=c_{1}(1,\tilde{\omega})-c_{0}(1,\text{\ensuremath{\tilde{\omega}}})\frac{c_{2}(2,\tilde{\omega})}{c_{1}(2,\tilde{\omega})-c_{0}(2,\text{\ensuremath{\tilde{\omega}}})\frac{....}{....}}....,\label{eq: continued fractions}
\end{equation}
which is a condition for the existence of the quasi-normal modes.

As $R_{m}\rightarrow-1\:$ for large $m\;$ and the changes of $R_{m}$
slow down for large $|\tilde{\omega}|$, assuming that $R_{m}$ changes
adiabatically is an excellent approximation and one gets \cite{key-16}
\begin{equation}
\frac{R_{m}}{R_{m-1}}\simeq1+\mathcal{O}(\tilde{\omega}^{-\frac{1}{2}}).\label{eq: ordine}
\end{equation}
This approximation works for both $Re\left(m+2i\text{\ensuremath{\tilde{\omega}}}\right)>0,$
when one computes $R_{m}$ recursively from $R_{\infty}=\text{\textminus}1$,
and for $Re\left(m+2i\text{\ensuremath{\tilde{\omega}}}\right)<0,$
when one starts to compute from $R_{0}=\text{\ensuremath{\infty} }$
\cite{key-16}. If one inserts $R_{m-1}=R_{m}\;$ in eq. (\ref{eq: 1-0}),
one gets a quadratic equation, having the solutions (at the leading
terms for large $m$) \cite{key-16}

\begin{equation}
R_{m}^{\pm}=\frac{-(m+2i\text{\ensuremath{\tilde{\omega}}})\pm\sqrt{2i\text{\ensuremath{\tilde{\omega}}}(m+2i\text{\ensuremath{\tilde{\omega}}})}}{m}+\mathcal{O}(m^{-\frac{1}{2}}).\label{eq: ordinata}
\end{equation}
The approximated solution (\ref{eq: ordinata}) can be carefully checked
by using Mathematica \cite{key-16}. The issue that $R_{m}$ must
satisfy eq. (\ref{eq: ordinata}) for one of the signs is a necessary
condition \cite{key-16}. One needs a more deep discussion in order
to see if that condition is also sufficient \cite{key-16}. When $Re\left(m+2i\text{\ensuremath{\tilde{\omega}}}\right)<0,$
the sign arises from the condition $R_{1}$ is small. Two terms in
eq. (\ref{eq: ordinata}) are approximately deleted. The sign for
$Re\left(m+2i\text{\ensuremath{\tilde{\omega}}}\right)>0$ arises
from the condition $|R_{m}|<1$ for very large $m$ \cite{key-16}. 

When $|\tilde{\omega}|$ is very large (but minor than the total mass
of the black hole), one chooses an integer $N\:$ such that \cite{key-16}

\begin{equation}
1\ll N\ll|\tilde{\omega}|.\label{eq: molto minore}
\end{equation}
For the values of $N\:$ in eq. (\ref{eq: molto minore}), eq. (\ref{eq: ordinata})
can be used to determine $R_{\left[-2i\tilde{\omega}\right]\pm N}\;$
\cite{key-16} and only the second term in the RHS of eq. (\ref{eq: ordinata})
results to be relevant (the symbol for the integer part $\left[-2i\tilde{\omega}\right]$
represents an arbitrary integer differing from $-2i\tilde{\omega}\;$
by a number much smaller than $N\;$ which is assumed to be even)
\cite{key-16}. Such a relevant term in eq. (\ref{eq: ordinata})
implies \cite{key-16}

\begin{equation}
R_{\left[-2i\tilde{\omega}\right]+x}\propto\pm\frac{i\sqrt{x}}{\sqrt{-2i\tilde{\omega}}}\;\; for\;\;1\ll x\ll|\tilde{\omega}|,\label{eq: for}
\end{equation}
while the first term in eq. (\ref{eq: ordinata}) is $\propto\frac{x}{\tilde{\omega}}\;$
and results subleading \cite{key-16}. Neglecting such a first term
is equivalent to neglecting $c_{1}(m,\tilde{\omega})$ in the original
equation (\ref{eq: 1-0}). In fact, this term is irrelevant for all
the $m\:$ for large $|\tilde{\omega}|$, with the possible exclusion
of some purely imaginary frequencies where $c_{0}(m)$ or $c_{2}(m)$
vanish \cite{key-16}. The ratio $R_{\left[-2i\tilde{\omega}\right]\pm N}\;$
is computed from eq. (\ref{eq: ordinata}) like the ratio of $\sqrt{x}\;$
for $x=\pm N$ \cite{key-16}

\begin{equation}
\frac{R_{\left[-2i\tilde{\omega}\right]+N}}{R_{\left[-2i\tilde{\omega}\right]-N}}=\pm i+\mathcal{O}(\tilde{\omega}^{-\frac{1}{2}}).\label{eq: ratio}
\end{equation}
The assumptions that permitted to obtain eq. (\ref{eq: ordinata})
break down when $|m+2i\text{\ensuremath{\tilde{\omega}}}|\sim1$\cite{key-16}.
In that case, the coefficients $c_{0}(m),\; c_{1}(m),\; c_{2}(m)$
contain terms of order $1$ that cannot be neglected \cite{key-16}.
They also strongly depend on $m$. The adiabatic approximation breaks
down in this region and the quantities $R_{\left[-2i\tilde{\omega}\right]\pm N}\;$
have to be related through the original continued fraction. Below,
we will calculate the continued fraction exactly in the limit of very
large $|\tilde{\omega}|$. The continued fraction will give the same
result of eq. (\ref{eq: ratio}) like the adiabatic argument. In fact,
the two solutions will eventually \textquotedblleft{}connect\textquotedblright{}
\cite{key-16}. Such a connection will release a non-trivial constraint
on $\tilde{\omega}$. 

As $R_{\left[-2i\tilde{\omega}\right]+x}\propto\frac{1}{\sqrt{\tilde{\omega}}}$,
$\; c_{1}(m)$ in the denominator of eq. (\ref{eq: 1-0 bis}) is negligible
when compared to the other term (which results $\propto\sqrt{\tilde{\omega}}$
) \cite{key-16}. By fixing $N$, one understands that the effect
of $c_{1}(m)$ in eq. (\ref{eq: 1-0 bis}) vanishes for large $|\tilde{\omega}|$\cite{key-16}.
An exception should appear when $c_{0}(m)$ and/or $c_{2}(m)$ $\rightarrow0$
for some $m,$ but we will show that this exception cannot occur when
$Re(\tilde{\omega})\neq0$. 

We note that the orbital angular momentum $l\:$ is irrelevant because
$c_{1}(m)$ does not affect the asymptotic frequencies \cite{key-16}.
This is not surprising as it is in agreement with Bohr correspondence
principle \cite{key-10,key-25,key-26,key-27}. One can replace the
factor $m$ in eq. (\ref{eq: c0}) with $\left[-2i\tilde{\omega}\right].$
In fact, $|\tilde{\omega}|$ becomes extremely large when one studies
only a relatively small neighborhood of $m\sim\left[-2i\tilde{\omega}\right]$
\cite{key-16}. We also see that the continued fraction is simplified
into an ordinary fraction. If one inserts eq. (\ref{eq: 1-0 bis})
recursively into itself $2N$ times, one gets \cite{key-16}

\begin{equation}
R_{\left[-2i\tilde{\omega}\right]-N}=\prod_{k=1}^{N}\left(\frac{c_{2}(\left[-2i\tilde{\omega}\right]-N+2k-1)c_{0}(\left[-2i\tilde{\omega}\right]-N+2k)}{c_{0}(\left[-2i\tilde{\omega}\right]-N+2k-1)c_{2}(\left[-2i\tilde{\omega}\right]-N+2k)}\right)R_{\left[-2i\tilde{\omega}\right]+N}.\label{eq: Produttore}
\end{equation}
The dependence of the $\; c_{k}(m)$ on the frequency is suppressed.
Thus, one can combine eqs. (\ref{eq: ratio}) and (\ref{eq: Produttore})
to eliminate the other $R_{m}$. We require that the generic solution
(\ref{eq: ordinata}), which holds almost everywhere, is \textquotedblleft{}patched\textquotedblright{}
with the solution (\ref{eq: Produttore}), which holds for $m\sim\left[-2i\tilde{\omega}\right]$
and for $|\tilde{\omega}|$ extremely large \cite{key-16}. 

One can express the products of $c_{0}(m)$ and $c_{2}(m)$ in terms
of the gamma function $\Gamma,$ which is an extension of the factorial
function to real and complex numbers \cite{key-16}. As $c_{2}(m)$
is bilinear, the four factors of eq. (\ref{eq: Produttore}) lead
to a product of $6$ factors, i.e. $(2+1+1+2).$ Each of those equals
a ratio of two $\Gamma$ functions. In other words, one gets a ratio
of twelve $\Gamma$ functions \cite{key-16}. 

One can write down the resulting condition in terms of a shifted frequency
defined by $-2if\equiv-2i\tilde{\omega}-\left[-2i\tilde{\omega}\right]$
\cite{key-16}:

\begin{equation}
\pm i=\begin{array}{c}
\frac{\Gamma\left(\frac{N+2if+j}{2}\right)\Gamma\left(\frac{N+2if-j}{2}\right)\Gamma\left(\frac{-N+2if+1}{2}\right)}{\Gamma\left(\frac{-N+2if+j}{2}\right)\Gamma\left(\frac{-N+2if-j}{2}\right)\Gamma\left(\frac{+N+2if+1}{2}\right)}\\
\\
\frac{\Gamma\left(\frac{-N+2if+j+1}{2}\right)\Gamma\left(\frac{-N+2if-j+1}{2}\right)\Gamma\left(\frac{+N+2if+2}{2}\right)}{\Gamma\left(\frac{N+2if+j+1}{2}\right)\Gamma\left(\frac{N+2if-j+1}{2}\right)\Gamma\left(\frac{-N+2if+2}{2}\right)}
\end{array}\label{eq: gamma-casini}
\end{equation}
Considering eq. (\ref{eq: c3}), the factors $m$ result cancelled.
Six of the $\Gamma$ functions in eq. (\ref{eq: gamma-casini}) show
an argument with a huge negative real part. Hence, they can be converted
into $\Gamma$ of positive numbers by using the formula \cite{key-16}

\begin{equation}
\Gamma(x)=\frac{\pi}{\sin\left(\pi x\right)\Gamma(1-x)}.\label{eq: gamma formula}
\end{equation}
Thus, the $\pi$ factors cancel like the Stirling approximations for
the $\;\Gamma$ functions which have a huge positive argument \cite{key-16}

\begin{equation}
\Gamma(m+1)\approx\sqrt{2\text{\ensuremath{\pi}}n}\left(\frac{m}{e}\right)^{m},\label{eq: Stirling}
\end{equation}
while the factors with $\sin\left(x\right)$ survive. The necessary
condition for regular frequencies for large $m\;$ (the frequencies
for which the analysis is valid. We will return on this point later)
reads \cite{key-16}

\begin{equation}
\pm i=\frac{\sin\left[\pi(if+1)\right]\sin\left[\pi(if+\frac{j}{2})\right]\sin\left[\pi(if-\frac{j}{2})\right]}{\sin\left[\pi(if+\frac{1}{2})\right]\sin\left[\pi(if+\frac{1+j}{2})\right]\sin\left[\pi(if+\frac{1-j}{2})\right]}.\label{eq: i}
\end{equation}
Choosing $N\in4\mathbb{Z},$ one erases $N\;$ from the arguments
of the trigonometric functions. One also replaces $f\:$ by $\tilde{\omega}$
again as the functions in eq. (\ref{eq: i}) are periodic with the
right periodicity and the number $\left[-2i\tilde{\omega}\right]$
can be chosen even. We can use the terms $\frac{\pi}{2}\;$ in the
denominator in order to convert the $\sin$ functions into $\cos$.
Multiplying eq. (\ref{eq: i}) by the denominator of the RHS and expanding
the $\sin$ functions in terms of the exponentials (one has to be
careful about the signs) the result is ($\epsilon(y)$ is the sign
function) \cite{key-16}

\begin{equation}
\exp\left[\epsilon Re\left(\omega\right)\cdot4\pi\tilde{\omega}\right]=-1-2\cos(\pi j).\label{eq: implica}
\end{equation}
Hence, for scalar or gravitational perturbations the allowed frequencies
are \cite{key-16} 
\begin{equation}
\tilde{\omega}_{m}=\frac{i}{2}(m-\frac{1}{2})\pm\frac{\ln3}{4\pi}+\mathcal{O}(m^{-\frac{1}{2}}),\label{eq: piuomeno rottura}
\end{equation}
while for vector perturbations one gets 
\begin{equation}
\tilde{\omega}_{m}=\frac{im}{2}+\mathcal{O}(m^{-\frac{1}{2}}).\label{eq: vettoriali}
\end{equation}
We note that, as we are in the large $m$ regime, one gets $\tilde{\omega}_{m}\simeq\frac{im}{2}\:$
independent of $j.$ This implies the correctness of eq. (\ref{eq: andamento asintotico})
and the important issue that the quantum of area (\ref{eq: 8 pi planck})
is an intrinsic property of Schwarzschild BHs. Again, this is in full
agreement with Bohr correspondence principle \cite{key-10,key-25,key-26,key-27}.

In order to finalize the analysis one has to resolve the question
marks concerning the special frequencies where the used approximation,
which neglects $c_{1}(m)$, breaks down \cite{key-16}.

The first step is to argue that the \textquotedblleft{}regular\textquotedblright{}
solutions must exist \cite{key-16}. In fact, if one can relate the
remainders $R_{\left[-2i\tilde{\omega}\right]\pm N}$ by using the
continued fraction, where $c_{1}(m)$ can be neglected, one can also
extrapolate them to eq. (\ref{eq: ordinata}) \cite{key-16}. 

From the boundary conditions $R_{0}=\infty$, $R_{\infty}=\lyxmathsym{\textminus}1^{+},$
one sees that a specific sign of the square root in eq. (\ref{eq: ordinata})
has to be separately chosen for $\; m<\left[-2i\tilde{\omega}\right]$
and for $\; m>\left[-2i\tilde{\omega}\right]$ \cite{key-16}. But
one finds that the signs agree with the signs of $\pm i\;$ that automatically
lead to the solutions \cite{key-16}. 

The condition for $\tilde{\omega}$ is both \emph{necessary and sufficient
}\cite{key-16}. This kind of solutions are only the $\ln(3$) solutions
from eq. (\ref{eq: rottura}). The existence of those solutions is
guaranteed \cite{key-16}. 

As one needs to find all irregular solutions, let us recall two useful
points \cite{key-16}. 
\begin{enumerate}
\item The continued fractions of eq. (\ref{eq: continued fractions}) depend
on the coefficients $c_{0}(m)$ and $c_{2}(m+1)$ \emph{only} through
their product $c_{0}(m)c_{2}(m+1)$ \cite{key-16}. 
\item If one finds zeroes in eq. (\ref{eq: Produttore}) exclusively in
the numerator or in the denominator, the ratio $\frac{R_{\left[-2i\tilde{\omega}\right]-N}}{R_{\left[-2i\tilde{\omega}\right]+N}}$
can be only either zero, or infinite. When one takes into account
$c_{1}(m)$, \textquotedblleft{}zero\textquotedblright{} or \textquotedblleft{}infinity\textquotedblright{}
results to be replaced by a negative or positive power of $|\tilde{\omega}|$,
respectively \cite{key-16}. 
\end{enumerate}
Point 2. means that one can obtain irregular solutions only by finding
$\omega$ such that eq. (\ref{eq: Produttore}) becomes an indeterminate
form $\frac{0}{0}$ \cite{key-16}. 

As $c_{0}(m)$ can be null at most for one value of $m$, one finds
that there is at least one value of $m$ where $c_{2}(m)$ vanishes
\cite{key-16}. Thus, eq. (\ref{eq: c3}) implies that one between
$2\left(i\tilde{\omega}+1\right)$ and $\;2\left(i\tilde{\omega}-1\right)$
(maybe both) must be integer \cite{key-16}. As the two conditions
are equivalent, both numbers $2\left(i\tilde{\omega}\pm1\right)$
must be integers to give a chance to exist to the quasi-normal frequency
\cite{key-16}. Thus, the two numbers differ by an even number. Then,
both the vanishing factors of $\; c_{2}(m)$ must appear in the numerator
of eq. (\ref{eq: Produttore}), or, alternatively,  both must appear
in the denominator of such an equation \cite{key-16}. One can assume,
for example, that they appear in the denominator without loss of generality
\cite{key-16}. Thus, one finds the indeterminate form only if the
vanishing $c_{0}(m)$ appears in the numerator \cite{key-16}. Clearly,
$2\left(i\tilde{\omega}\pm1\right)$ and $2i\tilde{\omega}$ are different
modulo two. Thus, the effect of $c_{1}(m)$ gives the desired result,
confirming that the regular states of eq. (\ref{eq: piuomeno rottura})
are the \emph{only} solutions \cite{key-16}. By using eqs. (\ref{eq: important definition})
and (\ref{eq: set}) one easily returns to Planck units and obtains
eq. (\ref{eq: quasinormal modes corrected}).

Now, we show that the continually rescaled mass unit in the above
discussion, which is due to the frequency dependence of $M_{E}$ and
$r_{E}$, did not influence the analysis. We note that, although $\tilde{\omega}$
in the analysis can be very large because of definition (\ref{eq: important definition}),
$\omega$ must instead be always minor than the BH initial mass as
BHs cannot emit more energy than their total mass. Inserting this
constrain in eq. (\ref{eq: effective quantities}) we obtain the range
of permitted values of $M_{E}(|\omega_{n}|)$ as 

\begin{equation}
\frac{M}{2}\leq M_{E}(|\omega_{n}|)\leq M.\label{eq: range 1}
\end{equation}
Thus, setting $2M_{E}(|\omega_{n}|)=r_{E}(|\omega_{n}|)\equiv1(|\omega_{n}|)$
one sees that the range of permitted values of the continually rescaled
mass unit is always included within a factor $2.$ On the other hand,
we recall that the countable sequence of QNMs is very large, see Section
3 and \cite{key-4}. Thus, the mass unit's rescaling is extremely
slow. Hence, one can easily check, by reviewing the above discussion
step by step, that the continually rescaled mass unit did not influence
the analysis.

Another argument which remarks the correctness of the analysis in
this Appendix is the following. One can choose to consider $M_{E}$
as being constant within the range (\ref{eq: range 1}). In that case,
it is easy to show that such an approximation is indeed very good.
In fact, eq. (\ref{eq: range 1}) implies that the range of permitted
values of $T_{E}(|\omega_{n}|)$ is 
\begin{equation}
T_{H}=T_{E}(0)\leq T_{E}(|\omega_{n}|)\leq2T_{H}=T_{E}(|\omega_{n_{max}}|),\label{eq: range 2}
\end{equation}
where $T_{H}$ is the initial BH Hawking temperature. Therefore, if
one fixes $M_{E}=\frac{M}{2}$ in the analysis, the approximate result
is 

\begin{equation}
\omega_{n}\simeq2\pi in\times2T_{H}.\label{eq: approssima 1}
\end{equation}
On the other hand, if one fixes $M_{E}=M\:$ (thermal approximation),
the approximate result is 

\begin{equation}
\omega_{n}\simeq2\pi in\times T_{H}.\label{eq: approssima 2}
\end{equation}
As both the approximate results in correspondence of the extreme values
in the range (\ref{eq: range 1}) have the same order of magnitude,
fixing $2M_{E}=r_{E}\equiv1$ does not change the order of magnitude
of the final (approximated) result with respect to the exact result.
In particular, if we set $T_{E}=\frac{3}{2}T_{H}$ the uncertainty
in the final result is $0.33$, while in the result of the thermal
approximation (\ref{eq: approssima 2}) the uncertainty is $2.$ Thus,
even considering $M_{E}$ as constant, our result is more precise
than the thermal approximation of previous literature and the order
of magnitude of the total emitted energies (\ref{eq: radice fisica})
is correct.

\end{document}